`

# Koza and Koza-Hub for born-interoperable knowledge graph generation using KGX

Daniel R Korn[1], Patrick Golden[1], Aaron Odell[1], Katherina Cortes[2], Shilpa Sundar[3], Kevin Schaper[1], Sarah Gehrke[1], Corey Cox[1], Harry Caufield[4], Justin Reese[4], Evan Morris[5], Christopher J Mungall[4], Melissa Haendel[1]

[1] TISLab, Department of Genetics, University of North Carolina at Chapel Hill
[2] Department of Biomedical Informatics, University of Colorado Anschutz Medical Campus
[3] Carolina Health Informatics Program, University of North Carolina at Chapel Hill
[4] Environmental Genomics and Systems Biology, Lawrence Berkeley National Laboratory
[5] Renaissance Computing Institute, University of North Carolina at Chapel Hill, Chapel Hill, NC, United States

## Abstract

Knowledge graph construction has become an essential domain for the future of biomedical research. But current approaches demand a high amount of redundant labor. These redundancies are the result of the lack of data standards and "knowledge-graph ready" data from sources. Using the KGX standard, we aim to solve these issues. Herein we introduce Koza and the Koza-Hub, a Python software package which streamlines ingesting raw biomedical information into the KGX format, and an associated set of conversion processes for thirty gold standard biomedical data sources. Our approach is to turn knowledge graph ingests into a set of primitive operations, provide configuration through YAML files, and enforce compliance with the chosen data schema.

`

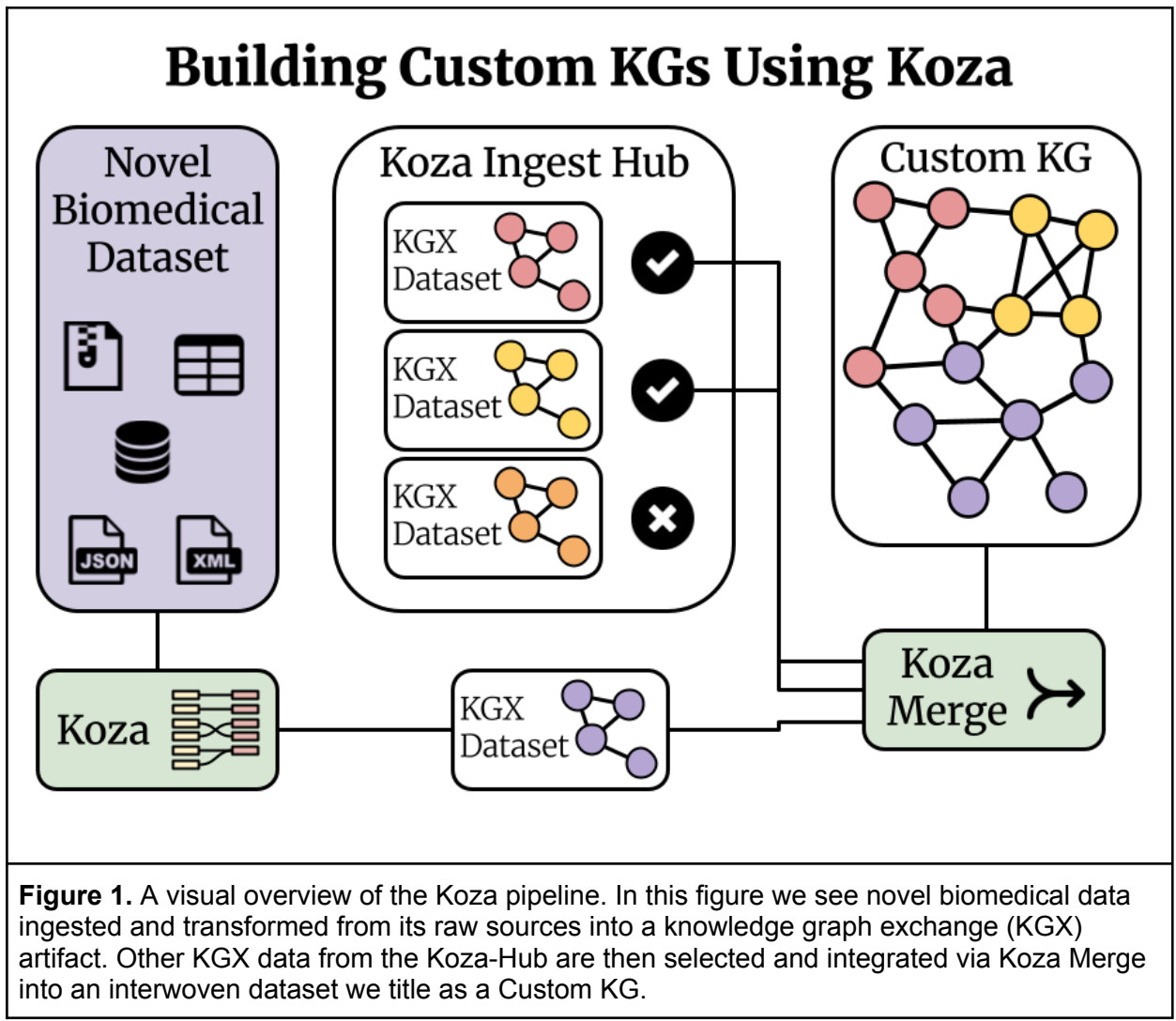

**Figure 1.** A visual overview of the Koza pipeline. In this figure we see novel biomedical data ingested and transformed from its raw sources into a knowledge graph exchange (KGX) artifact. Other KGX data from the Koza-Hub are then selected and integrated via Koza Merge into an interwoven dataset we title as a Custom KG.

# Introduction

The real power of data lies in how it fits into a larger context. For a long time, scientific knowledge has largely been confined to particular research domains[1,2]. Aiming to unite information across disciplinary boundaries into a centralized repository, biomedical Knowledge Graphs (KGs) provide researchers a way to interweave corpora produced across organizational and disciplinary boundaries. By relating knowledge across multiple domains and types of bioinformatics, we are able to reveal unsurfaced interrelationships between previously siloed data sources.

One key problem faced by KGs is the issue of consistent, repeatable, and transparent construction. Each time a new KG is constructed, there is a general trend towards recreation of entire build pipelines, starting from raw data and performing all the necessary parsing, formatting, interpretation, and transformations necessary to turn multiple non-KG data into a KG. Prominent KG construction surveys do not address the reusability of semi-structured data

`

[3]. Instead, KGs are often discussed as though they emerge ex nihilo, without consideration of potential data reuse. If we want to achieve the goal of "explainable AI", then the construction of the knowledge sources also need to be explained, what was excluded, and why, etc.

To address this gap, standards such as the Biolink Model have been proposed. The Biolink Model provides a high-level schema and controlled vocabulary for representing biomedical data and describing inter-domain relationships [4]. KGX is a data standard which utilizes Biolink to create an overarching, consistent, and computable model of the concepts and relationships at the heart of all of biomedical science. It is paired with the KGX toolkit, which is a reference implementation that can create serialized data objects that can be stored and shared (in this paper when we refer to KGX, we refer to the underlying data model and the reference implementation used in conjunction). KGX has been integrated into a variety of projects, including the Biomedical Data Translator project [5].

KGs are constructed from a variety of different sources transformed into a unified, connected data model, such that assertions can be connected across source boundaries. For example, a KG focused on disease pathways may combine multiple, disparate datasets about diseases, phenotypes, drugs, and genomes. The goal of a KG is to reconcile these disparate datasets such that the information contained in each is connected to the others. We refer to the process of transforming each dataset into the target data model an "*ingest*". Koza is a tool meant to facilitate and streamline the creation of these ingests .

To advance the reusability and standardization of KG construction, we introduce the Koza graph construction pipeline (see a visual illustration of the Koza pipeline in *Figure 1*). To our knowledge, the Koza pipeline is the first KG construction tool which combines: (1) modularity, (2) developer simplicity (here we consider anyone who may be a user of Koza a developer), (3) declarative parameters, (4) centering around a standard biomedical data standard (KGX), and (5) deploying as a standalone Python package. These choices enable (1) faster experimentation, (2) simplification of graph building, (3) making explicit previously implicit decisions, and (4) easy testability.

# Methodology

## Koza Primitives

Koza is fundamentally designed around the process of reading input data, transforming records into a target data model, and outputting those standard records into a standardized output file. These steps involve, respectively, three core components: a **reader**, a **writer, and** a **transform**.

**Readers** in Koza define how records are read into a pipeline. They are configured declaratively via sets of parameters defined in a YAML file which enable the ready parsing of structured data files that may be consumed. Most are specific to a certain file format: a CSV reader can be configured to interpret delimiter characters, comment characters, and header configuration, while a JSON reader can be configured to look for data in a certain path. All readers are able to

`

define filters in a small matching language to accept or reject records given certain criteria (e.g. membership in a list, comparison to a numeric value).

**Writers** in Koza, also configured via YAML parameters, control how records transformed from input data are converted to an output file. The writers that are currently part of Koza are built around the assumption that transforms will be creating property graphs, and output separate files for nodes and edges. Specifically, objects are expected to adhere to the KGX format.

**Transforms** are Python files that declare how a record should be transformed from input data to output data. To accomplish this, Koza uses a decorator-based API to declare that certain functions should be used to transform records. We refer to these functions as "hooks". The two main hooks are `@transform_record` and `@transform`. Exactly one of these decorators must be used in a transform file to denote to Koza the function that should be used to consume data from the reader. The former decorates a function to be called for every single record from the reader, while the latter decorates a function to be called once with the expectation that iteration will be handled manually by the consumer. All function hooks are passed through a KozaTransform object as their first argument, which allows functions to communicate with the transform process. Among the properties this object has is `write` (to send Biolink records to the writer), `state` (to store state across a transform), and `log` (to log messages to stdout). In practice, the decorator-based API looks like this:

```
@koza.transform_record()
def process_line(kt: KozaTransform, record: dict[str, Any]):
    protein = Protein(id=record["ID"], label=record["LABEL"])
    kt.write(protein)
```

Additional hooks to control the transform process include `@prepare_data` to run preliminary data cleaning, conversion, or generation before the transform begins. Other hooks include `@on_data_begin` and `@on_data_end` to perform logging or other actions before or after reading has concluded.

## Modularity

While knowledge graphs vary in the format and semantics of the data model that they target, Koza has been designed towards the KGX format and the Biolink data model. KGX is a format which serializes a property graph, consisting of nodes (entities) and edges (associations between entities, with optional qualifiers). In KGX, nodes and edges are expected to conform to the Biolink model, a standardized schema that enables the expression of concepts found in biomedical datasets. Nodes are named things, for example the anatomic entity *kidney*, the disease *hemophilia* or the phenotypes a*bnormal white blood cell count* and *hyperglycinemia*. Edges are the relations that may exist between those entities, such as *causes* or *results in phenotype*. Biolink offers a standardized way to describe *virus*, *disease*, and *phenotype* entities, as well as associations like *disease-to-phenotype*. The goal of a Koza ingest is to transform some arbitrarily structured data (contained in, for example, a CSV published by a research organization) into the KGX data model such that its nodes and edges can be connected to other similarly transformed data.

`

Instead of transforming multiple datasets in a large monolithic process, Koza is best used to develop per-source ingests separately, which are then concatenated into a larger KG. We call these per-source transformations **modular ingests**. Focusing each ingest on a particular data source gives developers and curators the ability to document and test the curatorial choices made for the particular semantics of a given dataset within a single unit.

The process of creating modular ingests enables streamlining the computational pipeline that is performed during KG construction (downloading from source, loading data, parsing, creating structured outputs). Additionally, processing once and re-releasing prevents raw data sources from being over used for KG build processes, which can over time become substantial demands on the source data providers (for example: GWAS is a source data provider) that they may not be resourced to handle. This especially applies to more complex data loads, such as SQL databases. Since KGX files can always be iterated/streamed by a programming language in a line-by-line manner, memory requirements to read through them are always substantially lower, or are equivalent to working with original datasets.

**Table 1.** Table which categorizes all the current set of modular ingests. Many of these ingests provide processing for multiple artifacts for a singular data set.

| Ingest Name | Ingest URL | Number of Artifacts Processed |
|---|---|---|
| alliance_genotype | https://github.com/monarch-initiative/alliance-genotype-ingest | 4 |
| alliance_phenotype | https://github.com/monarch-initiative/alliance-phenotype-association-ingest | 1 |
| alliance_disease_association | https://github.com/monarch-initiative/alliance-disease-association-ingest | 1 |
| biogrid | https://github.com/monarch-initiative/biogrid-ingest | 1 |
| clingen_variant | https://github.com/monarch-initiative/clingen-ingest | 1 |
| clinvar_variant | https://github.com/monarch-initiative/clinvar-ingest | 1 |
| hpoa | https://github.com/monarch-initiative/monarch-phenotype-profile-ingest | 4 |
| maxo_annotation | https://github.com/monarch-initiative/maxo-annotation-ingest | 1 |
| ncbi_gene | https://github.com/monarch-initiative/ncbi-gene | 5 |
| upheno_phenotype_to_phenotype | https://github.com/monarch-initiative/upheno-cross-species-ingest | 2 |
| pantherdb_ortholog | https://github.com/monarch-initiative/pantherdb-orthologs-ingest | 1 |
| zfin_genotype_to_phenotype | https://github.com/monarch-initiative/zfin-genotype-to-phenotype-ingest | 1 |
| go_annotation | https://github.com/monarch-initiative/go-ingest | 1 |
| zfin_orthology | https://github.com/monarch-initiative/zfin-orthology-ingest | 1 |

For implementation, we presently have our modular repositories on Github (*Table 1*) and we execute monthly releases through Github Actions. This collection serves as a regularly updated repository of biomedical knowledge standardized into KGX using Koza. We refer to this collection as the **Koza-Hub**, an expanding set of biomedical information which can be trivially incorporated into future KG builds.

This modularity process is not without its downsides. Any changes in the underlying knowledge sources will only be reflected in a KGX artifact once the modular build process has been run again. Additionally, Github limits data artifacts created by Github Actions to a filesize of 2 GB, which restricts the set of modular ingests we can create to relatively small/refined datasets.

# Use cases

Our organization curates data and creates and releases the Monarch KG on a monthly basis for the biomedical community. The Monarch KG serves as a structured collection of over 40 biomedical datasets (https://monarchinitiative.org/kg/sources), made available to the biomedical community as raw data releases (https://monarchinitiative.org/kg/downloads, https://data.monarchinitiative.org/monarch-kg) and also through a graph database interface (https://neo4j.monarchinitiative.org/).

## Use case 1 - Ingesting ClinGen

The Clinical Genome Resource (ClinGen) is a common data repository for the curation and validation of human variant pathogenicity, monogenic disease relationships, gene dosage sensitivity, and actionable gene-disease interventions [6]. Using Koza, we can readily integrate any of the ClinGen datasets into a KG-ready form by selecting the relationship types of interest for modeling. In particular, known pathogenic variants are of critical importance for the diagnosis of an individual with a suspected underlying genetic condition. To this end, we model ClinGen's pathogenic variants (Biolink:SequenceVariant) along with the genes (Biolink:Gene) and associated diseases (Biolink:Disease) using a modular Koza data ingest. This allows us to download the latest release of ClinGen every month, and integrate select information into the larger Monarch KG. Furthermore, as we expand the scope of the Monarch KG, we can integrate more data relationships from ClinGen, by modifying our existing ingest or creating a new one altogether. The current ClinGen Ingest is available at: https://github.com/monarch-initiative/clingen-ingest/.

## Use case 2 - Model Organism KG

In an effort to elucidate the functioning of model organisms, our team constructed the MOKG (Model Organism KG). Leveraging Koza tools, our team was able to build a specialized KG which specifically contained a curated corpus of information on model organisms. This KG was created by merging knowledge from nine sources, namely- HGNC [7], HPOA, uPheno [8], PHENIO [9,10], PANTHER [11], PomBase [12], Dictybase [13], Alliance of Genome Resources [14], and NCBI's Gene database [15]. We were able to construct these from existing individual Koza ingests created for the Monarch KG and producing one additional modular ingest resource. Moreover, we were able to repurpose the existing pipelines and release tooling to produce a Neo4J release of the tool. The MOKG can be viewed at: [https://mokg.monarchinitiative.org](https://mokg.monarchinitiative.org).

## Discussion

Data transformation is one of the more common tasks confronting scientists whose work involves taking advantage of external data sources. Many tools exist to transform data from one form to another.

The first advantage of Koza lies in its formalization of typical processes involved in data wrangling. One common method of transformation involves writing bespoke scripts in a programming language like Python or Perl. These scripts must cover (to a greater or lesser extent): downloading data, reading data, building new data structures, performing quality control, writing data, managing file hierarchies, handling errors, and logging diagnostics. Different projects and laboratories may implement these features in varying ways as personnel change over time. However, maintaining such bespoke scripts can become increasingly problematic. Koza addresses this challenge by providing interfaces that enable researchers to express these tasks succinctly, using declarative configurations and idiomatic Python functions. By employing Koza, researchers work with a well-defined set of components for common data-transformation tasks, without needing to focus on the minutiae of plumbing together data pipelines.

The second advantage of Koza is its special focus on data harmonization within the biomedical domain. The KG community as a whole faces persistent challenges with portability and interoperability of their graphs [16]. Many of these issues stem from the bespoke practices by which organizations import and structure data. By providing an easy to use codebase and a growing repository of existing ingests for many of the widely used biomedical data sources, Koza offers the community a straightforward and reusable alternative.

We are also working on addressing the deployment of modular data artifacts larger than 2GB (which we described in the **Modularity** section). Inevitably, issues related to hosting costs and logistics must be resolved to make this feasible. However, because dataset creation represents the longest and most resource-intensive stage of a KG construction pipeline, pre-computing and distributing these ingests could yield substantial financial and temporal benefits. Finding a way to pre-compute these ingests and provide them to the research community would enable creation of larger and more robust customized KGs. For example, resources like gnomAD provide data volumes that far exceed the processing capacity of many systems (the total size of all artifacts in gnomAD v4.1.0 is 1.66 TB) [17,18]. We hope to eventually petition data providers who have the bandwidth and community interest to provide their data in KGX format; this would allow it to remain constantly up to date with their latest releases and drive further community knowledge of these data standards.

# Conclusion

The Koza graph build system is already in use by multiple groups. The NIH funded Biomedical Data Translator project [19], KG-Microbe [20], the Monarch KG, and the MOKG all leverage the Koza pipeline in the construction of their knowledge graphs. The pipeline has benefited massively from the community contributions and attention; and has helped these projects create KGs quickly and allow their information to feed into other projects which have yet to be begun.

The Koza-Hub currently integrates data from 25 data artifacts across 14 data repositories and continues to expand. We aim to standardize additional datasets into knowledge graph–ready artifacts, thereby facilitating broader use within the biomedical community.

`